\documentclass[12pt]{iopart}
\usepackage{iopams} 

\usepackage{amsthm}
\usepackage{t1enc}
\usepackage{textcomp}
\usepackage[figuresright]{rotating}
\usepackage{subfig}
\usepackage{float}
\usepackage{hyperref}
\usepackage{afterpage}
\usepackage{upgreek} % used for \upmu
\usepackage{booktabs}  % publication-quality tables

\def\vec#1{\ensuremath{\mathchoice
                     {\mbox{\boldmath$\displaystyle\mathbf{#1}$}}
                     {\mbox{\boldmath$\textstyle\mathbf{#1}$}}
                     {\mbox{\boldmath$\scriptstyle\mathbf{#1}$}}
                     {\mbox{\boldmath$\scriptscriptstyle\mathbf{#1}$}}}}

\def\tens#1{\relax\ifmmode\mathsf{#1}\else\textsf{#1}\fi}

\renewcommand\div{{\rm div}}

\newcommand{\Delete} [1]{}
\newcommand{\Insert} [1]{#1}
\DeclareMathAlphabet{\Ibb}{U}{msb}{m}{n}

\newcommand{\Bb}{{\boldsymbol{\mathnormal b}}}

\newcommand{\Bk}{{\boldsymbol{\mathnormal k}}}
\newcommand{\Bl}{{\vec{l}}}
\newcommand{\Bn}{{\ensuremath\vec{n}}}

\newcommand{\Bq}{{\boldsymbol{\mathnormal q}}}

\newcommand{\Bs}{{\boldsymbol{\mathnormal s}}}
\newcommand{\BA}{{\vec{A}}}

\newcommand{\BQ}{{\vec{Q}}}
\newcommand{\BI}{{\vec{I}}}
\newcommand{\superscr}[1]{\ensuremath{{}^{\mathrm{#1}}}}
\newcommand{\subscr  }[1]{\ensuremath{{}_{\mathrm{#1}}}}

\newcommand{\Balpha} {\ensuremath{\boldsymbol\alpha}}

\newcommand{\tauext} {\ensuremath{           \tau \subscr{ext}}}

\newcommand{\tauy}   {\ensuremath{           \tau \subscr{f}}}

\newcommand{\Brho}   {\ensuremath{\boldsymbol\varrho}}
\newcommand{\Bkappa} {\ensuremath{\boldsymbol\kappa}}

\newcommand{\rhot}   {\ensuremath{           \rho \subscr{t}}}
\newcommand{\qt}     {\ensuremath{                q \subscr{t}}}

\newcommand{\ttot}   {\ensuremath{t_{\rm tot}}}
\newcommand{\tildettot}   {\ensuremath{\tilde{t}\subscr{tot}}}
\newcommand{\tildetauext}{\ensuremath{\tilde{\tau}\superscr{ext}}}

\newcommand{\dx}{\ensuremath{\mathrm{d}x}}
\newcommand{\dy}{\ensuremath{\mathrm{d}y}}

\newcommand{\dV}{\ensuremath{\mathrm{d}V}}

\newcommand{\taue}{\tau_{\rm ext}}

\usepackage{graphicx}% Include figure files
\usepackage{subfig}

\newcommand{\picA}[2]{%
	\subfloat[#1]{%
		\includegraphics[width=0.31\textwidth, viewport=35 10 555 415, clip]{figures/#2}%
	}
}

\newcommand{\eqref}[1]{(\ref{#1})}

\newcommand{\figref}[1]{Fig.~\ref{#1}}
\newcommand{\Figref}[1]{Fig.~\ref{#1}}
\newcommand{\secref}[1]{Sect.~\ref{#1}}

\begin{document}

\title{Pattern formation in a minimal model of continuum dislocation plasticity}
\author{Stefan Sandfeld and Michael Zaiser}
\address{Institute for Materials Simulation, 
         Department of Materials Science,
         Friedrich-Alexander University Erlangen-N\"urnberg (FAU), 
         Dr.-Mack-Str. 77, 90762 F\"urth, Germany}
\ead{stefan.sandfeld@fau.de}
\date{\today}

  \begin{abstract} %\footnotesize
The spontaneous emergence of heterogeneous dislocation patterns is a conspicuous feature of plastic deformation and strain hardening of crystalline solids. Despite long-standing efforts in the materials science and physics of defect communities, there is no general consensus regarding the physical mechanism which leads to the formation of dislocation patterns. In order to establish the fundamental mechanism, we formulate an extremely simplified, minimal model to investigate the formation of patterns based on the continuum theory of fluxes of curved dislocations. We demonstrate that strain hardening as embodied in a Taylor-type dislocation density dependence of the flow stress, in conjunction with the structure of the kinematic equations that govern dislocation motion under the action of external stresses, is already sufficient for the formation of dislocation patterns that are consistent with the principle of similitude.
  \end{abstract}

\submitto{\MSMSE}

%% \linenumbers

\section{Introduction}
\label{intro}
Work hardening during plastic deformation of crystalline solids is associated with significant changes in dislocation microstructure. The increase in dislocation density on the specimen scale is accompanied by the quasi spontaneous emergence of regions of low dislocation density and clusters of high dislocation density which to a large extent persist upon unloading. These metastable structures are denoted as \emph{dislocation patterns}. Despite a significant degree of morphological variation depending on slip geometry and loading mode (e.g. cell \cite{Kawasaki1980183} or labyrinth structures \cite{Zhang2003}, dislocation accumulation in  veins \cite{Siu2013_PhilMag93} or walls \cite{Mughrabi1983_Acta}), these patterns are characterized by some fairly universal scaling relationships. These relationships are commonly referred to as 'law of similitude' or 'similitude principle'. They relate the characteristic length $\Lambda$ of deformation-induced dislocation patterns to the applied stress $\tauext$ at which they have formed, and to their average dislocation density $\rho$: The wavelength $\Lambda$ is proportional to the dislocation spacing and inversely proportional to the applied stress, $\Lambda = D/\sqrt{\rho} = D aGb/\tauext$ where $G$ is the shear modulus, $b$ is the modulus of the Burgers vector, $a \approx 0.3$ is the non-dimensional coefficient relating flow stress and dislocation density in the Taylor relationship $\tauext = a G b \sqrt{\rho}$, and the remaining parameter $D$ is typically of the order of $D = 10\ldots 20$. These relations have been observed to hold in a wide range of materials and over more than four orders of magnitude in scale \cite{Rudolph2005_CrystResTech}. For a general overview, see e.g. the topical review by Sauzay\&Kubin \cite{Sauzay2011_ProgMaterSci56}.

Already for more than half a century attempts have been made to model dislocation patterning, using a huge variety of different simulation methods and theoretical approaches (see e.g. \cite{LeSar2014_AnnuRev5} for an overview). Early models were often based on analogies with other physical problems including spinodal decomposition patterns \cite{Holt1970_JAP}, chemical patterning in reaction-diffusion systems \cite{Walgraef1985, Pontes20061486}, noise-induced phase transitions \cite{Haehner1996,Haehner1999}, or even used simple internal-energy minimization arguments \cite{Hansen1986_MSE}. Despite the large number of published models, no consensus regarding the key mechanisms of dislocation patterning has emerged. In fact, it is only too easy to criticize much of the published literature as being inconsistent with basic observations about dislocation behavior and dislocation patterning: Energy minimization approaches which either consider energetic quasi-equilibrium ("low energy dislocation structures") or patterning in the approach to thermal equilibrium (spinodal decomposition) are inconsistent with the fact that dislocation patterning is only observed in intrinsically non-equilibrium situations where dislocations are driven by external stress. Analogies with reaction-diffusion models are spurious because it is difficult to see how the directed motion of dislocations in response to Peach-Koehler forces could be described as a diffusion process. In our view many of the published approaches focus too strongly on the patterning aspect, i.e., they represent attempts at 'modeling patterns' using templates which do not fully account for the rather peculiar dynamic properties and interactions of dislocations.  Instead, we are convinced that the only viable method to arrive at a satisfactory theoretical description of dislocation patterns consists in 'modeling dislocations', i.e., developing models that are capable of describing the stress-driven motion and interactions of curved and flexible dislocation lines in crystal lattice structures subjected to specified external constraints: if we get the physics right, patterns are bound to emerge. 

The huge variety of modeling approaches is matched by a similarly wide range of opinions regarding the physical processes that drive dislocation patterning. The spinodal decomposition model of Holt \cite{Holt1970_JAP} considers reduction of the elastic energy stored in the long-range stress fields of dislocations to be the key driver. Kuhlmann-Wilsdorff and her school (see e.g. \cite{Hansen1986_MSE}) more generally hold that dislocation patterns represent the configurations of lowest internal energy that are kinematically accessible at a given stage of deformation. 
\Insert{By contrast, Mughrabi \cite{Mughrabi1983_Acta} points out that dislocation patterns evolve during deformation under the action of an applied stress and in conformity with constraints imposed by the deformation. In this picture, the heterogeneity in the dislocation distribution leads to heterogeneity in the local flow stress. The ensuing plastic strain heterogeneity needs to be elastically accommodated and causes substantial deformation-induced long-range internal stresses, accompanied by an appreciable stored elastic strain energy \cite{Mughrabi1988_RevPhysAppl23}.} %
Finally, Madec et al. \cite{Madec2002_Scripta} state that only short-range stresses (line tension), which control the formation of junctions, together with processes such as cross slip which help the entanglement of dislocations, are essential to patterning while long-range stress fields are irrelevant. 

To resolve these controversies it would be desirable to develop computational models which allow to activate or de-activate physical micro-mechanisms at will, and thus to decide which features of dislocation motion are essential and which ones are incidental to pattern formation - an approach chosen for instance by Madec et al. \cite{Madec2002_Scripta}. Three-dimensional discrete dislocation dynamics (3D DDD) simulations provide a complete description of dislocation motion and interactions on the level of individual dislocation lines and might therefore offer a suitable methodological framework for such an investigation. 
\Insert{Indeed, early stages of dislocation patterning such as the formation of irregular dislocation clusters, braids, or entanglements, have been frequently observed in simulations, see e.g. Madec et al. \cite{Madec2002_Scripta}, Devincre et al. \cite{Devincre2001_MaterSciEngA} and recently also Hussein et al. \cite{Hussein2015_ActaMater85}. No published investigation, however, allows to clearly identify the pattern wave length and to directly link this through the similitude scaling relation to the applied stress. The reason for this is simply the required size such a simulation would need to have. As observed by El-Awaady et~al. \cite{El-Awady2015_NatCommu6},  the deformation behavior does not depend on the absolute size of a (simulated or real) specimen but on the size in units of dislocation spacings -- a fact that is immediately obvious from the scaling properties of dislocation systems \cite{Zaiser2014_MSMSE}. What is a sufficient size for investigating dislocation patterns in a DDD simulation? Experimental dislocation patterns have typical wavelengths (spacings between cell walls or braids/veins) of the order of $D = 10...20$ dislocation spacings \cite{Rudolph2005_CrystResTech}. In order to clearly identify a pattern wavelength and to ensure that the pattern is representative of bulk behavior rather than governed by boundary effects, the system should exhibit regular features in all spatial directions, hence, its spatial dimension $L$ should at least be 3-4 times larger than the pattern wavelength which makes for at least a linear dimension $L \sqrt{\rho}$ of 30-40 dislocation spacings. This is at the upper end of the system sizes attainable by state-of-the art 3D DDD simulations. Table 1, which shows values of system size and final dislocation density for some simulations reported in the literature, makes this point explicit. In fact, to our knowledge, there is only a single example of a 3D DDD simulation in the published literature where the simulated system size reaches some tens of the final dislocation spacing \cite{Arsenlis2007_MSMSE} and even there, the achieved strain of < 2\% may have been insufficient to form fully developed dislocation patterns. In all other mentioned studies, tendencies towards dislocation clustering concomitant with the formation of dislocation-depleted regions can be observed but the scale of these features is comparable to the simulated volume and hence no definite conclusions on pattern morphology or pattern wavelength can be drawn.
\begin{table}
	\centering\footnotesize
\begin{tabular}{ c | c | c | c }
\toprule
 & \bf linear   & \bf final dislocation  & \bf effective size  \\
 \bf reference & \bf dimension $L$ & \bf density $\rho$ & \bf $L \sqrt{\rho}$ \\
\hline
Madec et al. \cite{Madec2002_Scripta} & 10 $\upmu$m & $2.0 \times 10^{12}$ m$^{-2}$ & 14 \\
Arsenlis et al. \cite{Arsenlis2007_MSMSE} & 5 $\upmu$m & $7.5\times 10^{13}$ m$^{-2}$ & 42 \\
Hussein et al. \cite{Hussein2015_ActaMater85} & 10 $\upmu$m & $3.0\times 10^{12}$ m$^{-2}$ & 17 \\
  \bottomrule
\end{tabular}
\caption{Effective sizes of some DDD simulations reported in the literature}
\end{table}
}

Patterning can be efficiently simulated using two-dimensional discrete dislocation dynamics (2D DDD) where dislocations are envisaged as straight parallel lines, but such models require phenomenological rules to represent genuinely three-dimensional processes associated with segment curvature, such as dislocation multiplication and junction formation. Attempts have been made to formulate and parameterize such rules using 3D simulations ('2.5D DDD'), which results in patterns that are consistent with the similitude principle \cite{Gomez2006_PRL}. However, it is clear that this approach owing to its geometrical restrictions cannot give access to dislocation patterns in general deformation geometries. 

We expect that the above mentioned limitations of 3D DDD, which mostly relate to the huge computational cost of the method, can be overcome by resorting to density-based continuum dislocation dynamics (CDD) frameworks. A suitable CDD framework must be capable of representing the kinematics of curved dislocations in a geometrically correct manner and should allow to include the essential features of dislocation interactions. Several attempts in this direction have been reported in the literature. Some studies focus on subgrain formation (e.g. \cite{Chen2013_IJP}) - a process associated with geometrically necessary dislocations only - and can therefore not capture the initial stages of dislocation patterning which are associated with the clustering of statistically stored dislocations of zero net Burgers vector. Other investigations focus on specific mechanisms such as the sweeping of narrow dislocation dipoles by curved dislocations as envisaged by Kratochvil et al. \cite{Kratochvil2003_PRB} which may only be relevant in specific deformation conditions such as fatigue. 

In the present manuscript we undertake first steps towards a generic model that captures the coupled dynamics of statistically stored and geometrically necessary dislocations while accounting for the specific kinematics of curved dislocation lines. To this end we use the kinematic framework of Hochrainer's CDD theory to represent  fluxes of curved dislocations in a continuum setting  \cite{Hochrainer2007_PhilMag,Sandfeld2010_JMR,Hochrainer2014_JMPS, Hochrainer2015_PhilMag}. We use this framework to investigate the following question: Given the kinematics, what are the minimal ingredients required for the emergence of dislocation patterns?  In \secref{sec:model} we formulate our kinematic CDD framework together with a minimal model of dislocation interactions and a specification of boundary conditions. Scaling the system reveals the structure of the equations and allows to analyze the critical conditions required for patterning in \secref{sec:analysis}. We then use numerical simulation to investigate the properties of the emergent patterns and their evolution with stress during strain hardening (\secref{sec:num}). We conclude with a discussion of the physical mechanism of dislocation patterning and point out perspectives for future work. 

\section{The Continuum Model}
\label{sec:model}

\subsection{Continuum dislocation dynamics theory}
As kinematic framework on which we base our numerical investigation we use the CDD theory formulated by Hochrainer and co-workers \cite{Sandfeld2010_JMR,Hochrainer2014_JMPS, Hochrainer2015_PhilMag} (in the spirit of the present work one of the authors has also used the more complex, so-called 'higher-dimensional CDD' theory in a simple '1.5D model' of dislocation patterning \cite{Sandfeld2015_MRS}). The field variables describing the dislocation microstructure on a given slip system with normal vector $\Bn$ and Burgers vector $\Bb = b \Bs$ are the total density $\rhot$, the dislocation density vector $\Brho$, and the curvature density $\qt$. The latter quantity can be envisaged as an averaged product of dislocation line density and line curvature; its integral $\int_V \qt \dV = 2\pi N_{\rm d}$ over a volume $V$ yields the number of dislocation loops $ N_{\rm d}$ contained in the volume, hence, $\qt$ can also be envisaged as a loop density. We define the GND density vector as $\Brho = - (1/b) \nabla_\Bn \gamma$ where $\gamma$ is the plastic slip on the considered slip system, and $\nabla_\Bn = \nabla\cdot(\BI - \Bn \otimes \Bn)$ is the Nabla operator in the slip plane where $\BI$ is the rank-2 unit tensor. We assume that dislocation motion occurs by crystallographic glide with average (scalar) velocity $v$. Hence, Orowan's equation \cite{Orowan1940} gives the evolution of $\gamma$ as $\partial_t\gamma=\rhot b v$.

We note that our definition of $\Brho$ differs from the definition of the dislocation density vector $\Bkappa$ in \cite{Sandfeld2010_JMR,Hochrainer2014_JMPS, Hochrainer2015_PhilMag}; both formulations are related by $\Brho = - \Bn\times\Bkappa$ where $\Bn$ is the slip plane normal vector. The modified definition leads to an equivalent, but more intuitive formulation of the kinematic equations which are now given by 
% [\varrho1, \varrho2] = [ kappa2, -kappa1] = \kappa\perp !!!!!
\begin{eqnarray}
	\label{eq:drhotdt}
	\partial_t\rhot &=&-\nabla_\Bn\cdot(v\Brho)+v\qt\\
	\label{eq:drhovecdt}
	\partial_t\Brho &=& -\nabla_\Bn(v\rhot)\\
	\label{eq:dqtdt}
	\partial_t\qt&=&-\nabla_\Bn\cdot\left( -v\BQ^{(1)} + \BA^{(2)}\cdot \nabla_\Bn v \right).
\end{eqnarray} 
These equations can be considered lowest-order terms of an alignment tensor expansion \cite{Hochrainer2015_PhilMag}. We close this expansion by expressing the curvature density vector $\BQ^{(1)}$ and the rank-2 alignment tensor $\BA^{(2)}$ in terms of the variables $\rhot,\qt$ and $\Brho$. 

The curvature density vector $\BQ^{(1)}$ in \eqref{eq:dqtdt} plays very much the same role as the dislocation density vector $\Brho$ in \eqref{eq:drhovecdt}. As pointed out by Hochrainer et al.  {\cite{Hochrainer2014_JMPS}, $\BQ^{(1)}$ can be approximated as the product of the dislocation density vector and the mean curvature $\qt/\rhot$, 
\begin{equation}\label{eq:closea1}
%  \BQ^{(1)} = -\Bkappa^\perp \frac{\qt}{\rhot}.
  \BQ^{(1)} = -\Brho \frac{\qt}{\rhot}.
\end{equation}
In the vicinity of a homogeneous and isotropic dislocation arrangement, this approximation becomes exact. 
The rank-2 alignment tensor $\BA^{(2)}$ has trace ${\rm Tr}\, \BA^{(2)} = \rhot$ and for an isotropic dislocation arrangement, $\BA^{(2)} = 0.5\rhot \BI$. The deviatoric part $\BA^{(2)}_{\rm dev}$ characterizes imbalances in the distribution of dislocation characters (e.g. predominance of edge vs. screw dislocations). For a weakly anisotropic arrangement containing a fraction of geometrically necessary excess dislocations, we follow the proposal of Monavari et al. \cite{Monavari2014_MRS}) and write 
\begin{equation}\label{eq:closea2}
	{\BA}^{(2)}_{\rm dev} = \frac{\rhot}{2} \Phi(\Brho^2/\rhot^2)\left(\BI - 2 {\Bl}_{\rho} \otimes {\Bl}_{\rho}\right)
\end{equation}
where ${\Bl}_{\rho}$ is the unit vector in direction of the GND vector (the slip gradient) and $\Phi$ is a function with the properties $\Phi(0) = 0$ and $\Phi(1) = 1$. The approximation for $\BA^{(2)}$ proposed by Hochrainer et al. \cite{Hochrainer2014_JMPS} is tantamount to setting $\Phi(u) = \sqrt{u}$ which appears inappropriate since it leads to a formulation which is non-analytic at $u=0$, i.e., in the vicinity of a homogeneous and isotropic dislocation arrangement. We therefore follow the suggestion of Monavari et al. \cite{Monavari2014_MRS} who from a Maximum Entropy argument deduce $\Phi(u) \approx u(1+u)/2$.

\subsection{Dislocation interactions and dislocation velocity}
\label{sec:eq_of_motion}
In a density-based model, dislocation interactions enter in different ways depending on their range. Interactions on scales above the characteristic dislocation spacing (in the following referred to as {\em long-range interactions}) can be represented in terms of eigenstresses associated with a heterogeneous distribution of the plastic strain (\cite{Sandfeld2013_MSMSE,El-Azab2007_PhilMag_p1201}) or, equivalently, with the presence of geometrically necessary dislocations (non-vanishing dislocation density vector $\Brho$). Interactions on scales below the characteristic dislocation spacing, on the other hand (in the following referred to as {\em short-range interactions}) cannot be described in this manner, for the obvious reason that the continuous dislocation density fields and the associated strain field cannot meaningfully represent heterogeneities below the scale of the single-dislocation spacing. These short-range interactions are described in the standard manner in terms of a friction-like yield stress, which we assume to depend on the local value of the dislocation density according to the Taylor relationship.
Additional terms relating to line curvature or to short-range interactions between geometrically necessary dislocations might be considered (see the discussion of such terms by Zaiser \& Sandfeld \cite{Zaiser2014_MSMSE}) but,  in the spirit of formulating a minimal model, here we exclusively use the Taylor relationship to describe the short-range dislocation interactions which control the flow stress. The dislocation velocity is then computed from the glide component of the Peach-Koehler force as
\begin{equation} \label{eq:v}
v = \left\{
\begin{array}{cc}
(b/B)\mathrm{sign}(\tau)(|\tau|-\tauy) & \quad\textrm{IF}\quad |\tau|\geq \tauy, \\ 
0                                       & \quad\textrm{ELSE},
\end{array}\right.
\end{equation}
where $B$ is the drag coefficient. In this expression, $\tau$ is the resolved shear stress in the slip system, computed by solving the elastic boundary value problem for a body containing the eigenstrain field $\gamma$. This stress captures both the effect of the externally applied boundary tractions and/or constraints, and the long-range dislocation interactions associated with a heterogeneous plastic strain or, equivalently, with the dislocation field $\Brho$.  

\subsection{Boundary conditions}

We study two versions of the model, namely (1) a situation where long-range interactions are absent but short-range interactions are present, and (2) a situation with long-range interactions but no short-range ones. In both cases we envisage a layer of plastically active material of thickness $h$ where deformation occurs in single slip (a slip zone or slip band). The layer is assumed parallel to the $xy$ plane which is also the slip plane, and the slip direction is the $x$ direction. For numerical simulations, periodic boundary conditions are assumed in the $x$ and $y$ directions. 

In case (1) we assume that the plastically active layer forms a thin film loaded by homogeneous shear tractions acting on its unconstrained top and bottom surfaces. In case (2) we consider the layer to be embedded into an infinite block of elastic material which is loaded remotely, again in a spatially homogeneous manner. The stress field in the plastically active layer is then given by 
\begin{eqnarray}
\label{eq:tau}
\tau =  \taue,&  &{\rm case\, (1)},\nonumber\\
\tau =  \taue& + \int {\cal G}(x-x',y-y') \gamma(x',y')\, \dx'\,\dy',\qquad &{\rm case\, (2)},
\end{eqnarray}
where $\taue$ is the shear stress due to the externally applied tractions and the Fourier transform of the kernel $\cal G$ is given by (see e.g. \cite{Koslowski2002_JMPS})
\begin{equation}
\label{eq:g}
{\cal G}(k_x,k_y) = -\frac{Gh}{2}\left[\frac{k_y^2}{\sqrt{k_x^2+k_y^2}}+\frac{1}{1-\nu}\frac{k_x^2}{\sqrt{k_x^2+k_y^2}}\right].
\end{equation}
The convolution integral in \eqref{eq:tau} can from this relation be evaluated in standard manner by a double Fast Fourier Transform. 

\section{Mathenmatical analysis}
\label{sec:analysis}

\subsection{Dimensionless scaling }
\label{sec:scaling}
To analyze the behavior of the model it is convenient to switch to a non-dimensional representation which allows to identify the independent parameters. We define the scaling relations between quantities with physical units and their dimensionless counterparts (indicated by a tilde) as $\tau  =C_\tau \tilde\tau$ (for stresses), $\rhot  =C_\rho \tilde\rhot$ (for dislocation densities) and $x =C_x \tilde x$ (for lengths), with the scaling factors  
\begin{eqnarray}\label{eq:scaling_factors1}
	C_\tau= \taue, \qquad 
	C_\rho=\rho_0, \qquad
	C_x=\rho_0^{-1/2},
\end{eqnarray}
where $\rho_0 = \langle\rhot(t=0)\rangle$ is the mean initial dislocation density. The same scaling applies to the total dislocation density and the dislocation density vector. The scaling relation for the lengths implicates that lengths are measured in multiples of average dislocation spacings. For the velocity we find for $\tau > \tauy$ 
\begin{eqnarray}
&& \tilde{v} = v\frac{B}{b C_\tau} = 1  - {\cal H}_0\sqrt{\tilde{\rhot}} ,
\label{eq:v_scaled1}  
\end{eqnarray}
where ${\cal H}_0 = \alpha G b \sqrt{\rho_0}/\taue$. 

The scaling of velocity implies a scaling of time, since the scaling factor of velocity can be understood as the ratio between the characteristic length $C_x$ and a characteristic time $C_t$. Furthermore, the fact that the dislocation curvature has the dimension of an inverse length suggests to scale $\qt$ in proportion with $C_{\rho}/C_x$. Finally, we observe that the plastic slip can be envisaged as the product of the Burgers vector and the area per unit volume swept by dislocations. These observations lead to the derivative scaling relations $v = C_v \tilde{v}$, $t = C_t \tilde{t}$,  $\qt = C_q \tilde{\qt}$, and $\gamma = C_{\gamma} \tilde{\gamma}$ with
\begin{eqnarray}\label{eq:scaling_factors2}
C_v =  \frac{b}{B\taue},\quad
C_t = \frac{C_x}{C_v} = \frac{B\taue}{b\sqrt{\rho_0}},\quad
C_q = \frac{C_{\rho}}{C_x} = {\rho_0^{3/2}},\quad
C_{\gamma} = b \sqrt{\rho_0}.
\end{eqnarray}
An important implication of the scaling relationships for stress, length and dislocation density is that any patterns that may emerge from a simulation based on the scaled equations automatically fulfill the similitude principle (cf. \cite{Zaiser2014_MSMSE}) - provided that it can be demonstrated that the patterns do not depend on initial and boundary conditions (i.e., on system size and deformation history). 

In the scaled variables the kinetic equations read:
\begin{eqnarray}
	\label{eq:drhotdtt}
	\partial_{\tilde{t}}\tilde{\rho}_{\rm t} &=&-\tilde{\nabla}_\Bn\cdot(\tilde{v}\tilde{\Brho})+\tilde{v}\tilde{\qt},\\
	\label{eq:drhovecdtt}
	\partial_{\tilde{t}}\tilde{\Brho} &=& -\tilde{\nabla}_\Bn(\tilde{v}\tilde{\rhot}),\\
	\label{eq:dqtdtt}
	\partial_{\tilde{t}}\tilde{q}_{\rm t} &=&-\tilde{\nabla}_\Bn\cdot\left( -\tilde{v}\tilde{\BQ}^{(1)} + \tilde{\BA}^{(2)}\cdot \tilde{\nabla}_\Bn \tilde{v} \right),
\end{eqnarray} 
where the velocity $\tilde{v}$ is given by \eqref{eq:v_scaled1} and the quantities $\tilde{\BQ}^{(1)}$ and $\tilde{\BA}^{(2)}$ are obtained by replacing $\rho,\qt$ and $\Brho$ in \eqref{eq:closea1} with their scaled counterparts. The scaled stress is given by 
\begin{eqnarray}
\label{eq:taut}
\tilde{\tau} = 1,& &{\rm case\, (1)},\nonumber\\
\tilde{\tau} = 1& + \int \tilde{\cal G}(\tilde{x}-\tilde{x}',\tilde{y}-\tilde{y}') \tilde{\gamma}(\tilde{x}',\tilde{y}') \tilde{x}'\tilde{y}',\qquad &{\rm case\, (2)},
\end{eqnarray}
where \Insert{the Fourier representation of the kernel $\cal G$ is given by}
\begin{equation}
\tilde{\cal G}(\tilde{k}_x,\tilde{k}_y) = - {\cal G}_0 \left[\frac{\tilde{k}_y^2}{\tilde{k}}+\frac{1}{1-\nu}\frac{\tilde{k}_x}{\tilde{k}}\right]
\end{equation}
where $\tilde{k}=\sqrt{\tilde{k}_x^2+\tilde{k}_y^2}$ \Insert{and $\tilde{k}_x$, $\tilde{k}_y$ are the components of the wave vector in the $xy$-plane.}
The dislocation density vector relates to the scaled strain by $\tilde{\Bq} = - \tilde{\nabla}_n \tilde{\gamma}$, and the evolution equation for $\tilde{\gamma}$ is 
\begin{eqnarray}\label{eq:scaled_Orowan}
\partial_{\tilde{t}} \tilde{\gamma} = \tilde{\rhot}\tilde{v}.
\end{eqnarray}

It is interesting to note that, in the scaled formulation, the model has only two non-dimensional parameters. The first, ${\cal H}_0 = a G b \sqrt{\rho_0}/\taue$, measures the fraction of the externally applied stress that is needed to overcome short-range dislocation interactions. The second parameter, ${\cal G}_0 = G b h \rho_0/\taue$ measures the relative magnitude of the long-range dislocation interactions compared to the external stress. 

\subsection{Linear stability analysis}

\Insert{
To analyze  stability of the model, we envisage a spatially homogeneous reference state $\tilde{\rho}_h, \tilde{q}_h, \tilde{\gamma}_h$ characterized by the initial values $\tilde{\rho}_h = 1, \tilde{q}_h =  \tilde{\qt}^0$, $\tilde{\gamma}_h = \tilde{\gamma_0}$, with dislocation velocity $\tilde{v}_h = 1 - {\cal H}$ where ${\cal H} = {\cal H}_0 \sqrt{\tilde{\rho}_h}$ and with time evolution 
\begin{equation}
\partial_t \tilde{\gamma}_h = \tilde{\rho}_h \tilde{v}_h \:,\quad
\partial_t \tilde{\rho}_h =  \tilde{q}_h \tilde{v}_h\;,\qquad
\partial_t \tilde{q}_h = 0\:.
\end{equation}
We now study the evolution of small fluctuations around this reference state. We write these fluctuations as
\begin{equation}
\tilde{\gamma} = \tilde{\gamma}_h + \delta\tilde{\gamma}\;,\qquad
\tilde{\rhot} =  \tilde{\rho}_h + \delta\tilde{\rhot}\;,\qquad
\tilde{\qt} = \tilde{\qt}_0 - \frac{1}{2} \tilde{\Delta}_\Bn(\delta\tilde{\gamma}) + \delta \tilde{\qt}.
\end{equation}
}
Here, the operator $\tilde{\Delta}_\Bn = \tilde{\nabla}_\Bn \cdot \tilde{\nabla}_\Bn$ is the 2D Laplacian in the slip plane, and the corresponding term in the expression for the curvature density can be envisaged as the curvature of the geometrically necessary dislocations. In linear approximation, we find that 
\begin{equation}
\tilde{\BQ}^{(1)} =  \tilde{\qt}^0 \tilde{\nabla}_\Bn\delta\tilde{\gamma}\;,\qquad
\tilde{\BA}^{(2)} = \frac{1}{2} (1 + \delta\tilde{\rhot})\BI. 
\end{equation}
The linearized evolution equations for $\delta\tilde{\rhot}, \delta\tilde{\qt}$, and $\delta\tilde{\gamma}$ are then given by
\begin{equation}
\fl\partial_{\tilde{t}}\!
\left[\!\!\!\begin{array}{c}
\delta\tilde{\gamma}\\
\delta\tilde{\rhot}\\
\delta\tilde{\qt}
\end{array}\!\!\right]
\!=\!
\left[\!\!\begin{array}{ccc}
- \tilde{\cal G}_0 \left[\frac{\tilde{k}_y^2}{\tilde{k}}+\frac{1}{1-\nu}\frac{\tilde{k}_x^2}{\tilde{k}}\right]%\tilde{\cal G}(\tilde{k}_x,\tilde{k}_y) 
& 1-\frac{3 {\cal H}}{2}
& 0\\
-\frac{1+{\cal H}}{2} \tilde{k}^2 - \tilde{\qt}^0
\tilde{\cal G}_0 \left[\frac{\tilde{k}_y^2}{\tilde{k}}+\frac{1}{1-\nu}\frac{\tilde{k}}{|\tilde{\Bk}|}\right]
%\tilde{\cal G}(\tilde{k}_x,\tilde{k}_y) 
& - \frac{\tilde{\qt}^0{\cal H}}{2}               
& 1-{\cal H}\\
0 
& ({\cal H}-1)\tilde{k}^2 
& 0
\end{array}\!\!\right]\!\!
\left[\!\!\begin{array}{c}
\delta\tilde{\gamma}\\
\delta\tilde{\rhot}\\
\delta\tilde{\qt}
\end{array}\!\!\right].
\end{equation}
Eigenvalues of this matrix are $\Lambda=\Lambda(\tilde{\qt}^0, {\cal H}, {\cal G}_0, \tilde{\Bk})$. We first consider the hypothetical case where long-range stresses are appreciable but short-range stresses are absent (${\cal H} = 0,\; {\cal G}_0 > 0$). From the Routh-Hurwitz criterion it follows in this case that for any values of $\tilde{\Bk}$, all Eigenvalues have negative real part. The system is thus absolutely stable and no patterning can occur. More interesting is the opposite case, ${\cal G}_0 = 0, {\cal H} > 0$. In this case the Routh-Hurwitz criterion yields the condition 
\Insert{
$1 > {\cal H} > {\cal H}_1$, where ${\cal H}_1 = (9 - \sqrt{57})/2 \approx 0.725$, for eigenvalues with positive real part to occur.
} 
Within this unstable regime, the critical eigenvalue branch is real for all $\tilde k > 0$, indicating that fluctuations on all scales become undamped while only the homogeneous state $\tilde k = 0$ remains marginally stable.  Finally, if both ${\cal H}_0 > 0$ and ${\cal G}_0 > 0$, the instability condition remains unchanged, however, in this case long-wavelength fluctuations are damped and only fluctuations above a critical value $k_c$ can become unstable. The regime of unstable $k$ vectors is given by 
\begin{equation}
\tilde{k} > k_{\rm c}(\theta) = \frac{{\cal G}_0 [1 + {\cal H}_0] \tilde{\qt}^0}{({\cal H}_2 - {\cal H})({\cal H} - {\cal H}_1)} \left[\sin \theta + \frac{\cos \theta}{1 - \nu}\right] \label{Eq:kc}
\end{equation}
where ${\cal H}_{1/2} = (9 \pm \sqrt{57})/2$.
Examples showing the critical eigenvalue branch as a function of the wave-vector are shown in \figref{fig:eigenvalue}.

\begin{figure}[htbp]
\centering
\includegraphics[width=\textwidth]{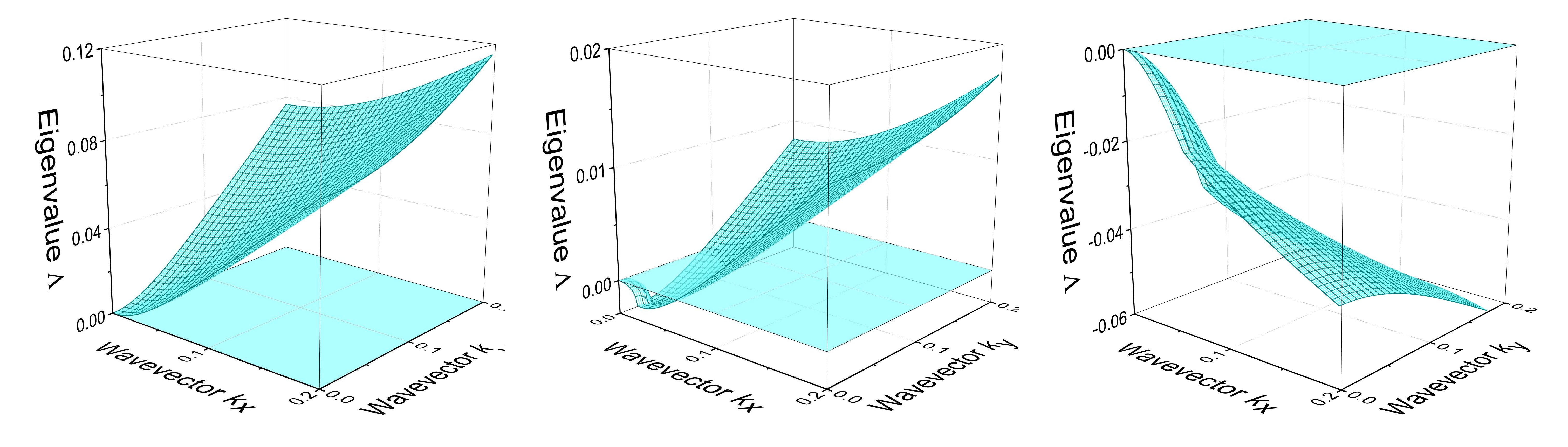}
\caption{Critical eigenvalue branch as a function of $\tilde{\Bk}=(\tilde{k}_x,\tilde{k}_y)$; initial curvature $\tilde{\qt}^0 = 0.1$; 
left: only short-range interactions (${\cal H} = 0.85, {\cal G} = 0$), center: long- and short-range interactions (${\cal H} = 0.85, {\cal G} = 1$), right: only long-range interactions (${\cal H} = 0, {\cal G} = 1$).} 
\label{fig:eigenvalue}
\end{figure}

\Insert{
Since the homogeneous reference state is time dependent with increasing dislocation density, one may ask whether instability in the above sense is a sufficient criterion for the emergence of heterogeneous dislocation patterns -- density fluctuations which grow more slowly than the statistically homogeneous reference density might in practice hardly be visible. While the ultimate verdict about patterning must consider the non-linear dynamics of the system and hence be provided by numerical simulation, one can take the growth of the reference density into consideration in a linear stability analysis. To this end one simply considers, instead of the absolute dislocation density fluctuation $\delta\rhot$, the relative fluctuation $\delta \rhot/\rho_h$. It turns out that this modification leaves the instability condition $1 > {\cal H}_0 > {\cal H}_1$ unchanged, but reduces the apparent growth rate of the patterns and decreases the unstable range of wavevectors for ${\cal G} > 0$ as the critical wavevector in Eq. (\ref{Eq:kc}) is increased by a factor of 2. 
}

In all cases where unstable modes occur, the unstable mode is characterized by antiphase fluctuations of $\gamma$ and $\rhot$ (regions of high dislocation density correspond to regions of low strain and vice versa). The same is true for the fluctuations of $\qt$ and $\rhot$, hence, the dislocation curvature (and thus the multiplication of dislocations) decreases in regions of high dislocation density and increases in regions of low density. This is consistent with the well known fact that, in heterogeneous dislocation patterns, multiplication occurs by the expansion of segments in the dislocation-depleted regions (channels or cell interiors). 

\section{Simulation of pattern formation}
\label{sec:num}

The analysis of the previous section can establish conditions for a homogeneous dislocation arrangement to become unstable. However, it cannot give access to the morphology of the emerging patterns which depends on the non-linear dynamics of the system. Moreover, the instability is in linear approximation not characterized by any dominant wavelength, hence, also the characteristic length of the patterns remains undetermined by the linear stability analysis. To access pattern morphology and pattern scale (if any), we resort to numerical simulation of the model. 

\subsection{Numerical methods}
For the numerical solution of the scaled CDD evolution equations \eqref{eq:drhotdtt}-\eqref{eq:dqtdtt}, \eqref{eq:scaled_Orowan} we use a Galerkin finite element scheme together with an implicit time integration. The spatial discretization consists of approximately 50 triangular elements in each spatial direction. In order to properly represent first and second derivatives within the finite element scheme we use third order Lagrangian shape functions. For reasons of numerical stability a small viscous term was added to each of the evolution equations, e.g. in \eqref{eq:drhotdtt} we replace the divergence term
$\tilde{\nabla}_\Bn\cdot(\tilde{v}\tilde{\Brho})$ % $\div(v\Brho)$ 
with 
$\tilde{\nabla}_\Bn\cdot(\tilde{v}\tilde{\Brho}  + \epsilon\nabla\tilde{\Brho})$, %$\div(v\Brho + \epsilon_\rho\nabla\rhot)$, 
where $\epsilon$ is chosen sufficiently small such that the physical behavior of the equations is not affected but numerical oscillations are suppressed. For each time step we update the velocity field (i.e. the Taylor term) and then evolve the dislocation fields. One of the benefits of the dimensionless scaling is that we can solve systems with widely different parameters based on the same finite element mesh; the small but artificial viscous terms for stabilizing the numerical scheme also have the same influence on the results for widely different systems, which improves our control over the numerics.

\subsection{Initial values}
Initial values for the dislocation fields should be consistent with representing a system of curved and connected lines in a coarse-grained manner. To this end we proceed in analogy with the coarse graining of a stochastic point process into a superposition of Gaussian 'blobs': we consider a system of discrete dislocation loops of radius $r$ and 'smear out' each loop perpendicular to its line tangent by using a Gaussian standard distribution function $(s \sqrt{2\pi})^{-1} \exp(-\xi^2/{2s^2})$  where the standard deviation $s$ characterizes the width of the distribution and $\xi$ is the distance perpendicular to the loop. After having obtained $\rhot^l$ for one loop in this way we determine the corresponding dislocation density vector $\Brho^l$ by multiplication of $\rhot^l$ with the radial unit vector, and the curvature density by division of $\rhot$ by $r$. Initial values for the fields are then obtained by randomly distributing such loops and summing up the loop fields $\rhot^l, \Brho^l$, and $\qt^l$. An example of a resulting initial structure is shown in \figref{fig:pattern}. An alternative method to define initial conditions is to define random scalar fields $\gamma$, $\qt$ and $\rhot$ as superpositions of randomly located Gaussian 'blobs' and determine $\Brho$ as the gradient of $\gamma$, which automatically fulfils the solenoidality condition $\div \Balpha$ of the Kr\"oner-Nye tensor. To make sure that our results do not depend on the construction of initial conditions, both methods have been used in the following simulations. 

\subsection{Parameter values}
We consider the system which is defined by the parameters given in table~\ref{table:parameter}. The scaled quantities were obtained from the physical values according to Eqs. \eqref{eq:scaling_factors1}, \eqref{eq:scaling_factors2}.
\begin{table}[htb] 
\centering \footnotesize 
$\begin{array}{l@{\quad}| r@{\,} l@{\,}l@{\quad}|r@{\,}l@{\qquad}} 
\toprule%\hline\hline 
                             &\multicolumn{3}{c|}{\textrm{\bf physical value}} &\multicolumn{2}{c}{\textrm{\bf scaled value}}\\ \hline 
\textrm{\bf Model parameter} &         &                    &             &                &        \\                              
\textrm{system size}         & l=      &  2.00\cdot 10^{-6}  &\textrm{m}   & \tilde l=      & 17.72   \\                          
\textrm{system height}       & h=      &  0.20\cdot 10^{-6}  &\textrm{m}   & \tilde h=      &  1.77   \\                          
\textrm{Taylor factor}       & a= 		 &  0.30               &\textrm{ }   & -              &        \\                            
\textrm{applied stress}      & \tauext=&  0.17\cdot 10^{9}   &\textrm{Pa}  & \tildetauext = &  1     \\                         
\textrm{total time}          & \ttot=  &  0.05\cdot 10^{-6}  &\textrm{s}   & \tildettot =   & 18.97   \\\hline                  
\textrm{\bf Material parameter} &      &                    &             &                &        \\                              
\textrm{Burgers vector modulus} & b=      & 0.256\cdot 10^{-9} &\textrm{m}   & \tilde b=      &  2.27\cdot 10^{-3}\\                
\textrm{Drag coefficient}    & B=      &  5.00\cdot 10^{-5}  &\textrm{Pa}\cdot\textrm{s} & \tilde B= &  2.27\cdot 10^{-3} \\   
\textrm{Shear modulus}       & G=      &   128\cdot 10^{9}  &\textrm{Pa}  & \tilde G=      &   1469  \\ \hline                  
\textrm{\bf Initial values}  &         &                    &             &                &        \\                              
\textrm{loop radius}         & r=      &  0.20\cdot 10^{-6}  &\textrm{m}   & \tilde r=      & 1.772   \\                          
\textrm{number of loops}     & N=      &    50                 &\textrm{ }   & -              &        \\                              
\textrm{average density}     & \rho_0=  &  7.85 \cdot 10^{13} &\textrm{ }  & \tilde{\rho}_0=    & 1\\                         
\textrm{standard deviation}  & s=      &    25 \cdot 10^{-9}   &\textrm{m}   & \tilde s=      & 0.222    \\                              
\bottomrule%\hline\hline                              
\end{array} $                              
\caption{\label{table:parameter} System and material parameters used for the simulation shown in \figref{fig:pattern}. Dimensionless (scaled) parameters and values are indicated by a tilde.}  
\end{table} 

\begin{figure}[p]
\centering\footnotesize
{\bf Initial Values}\\[-0.6em]
\picA{$\rhot$}{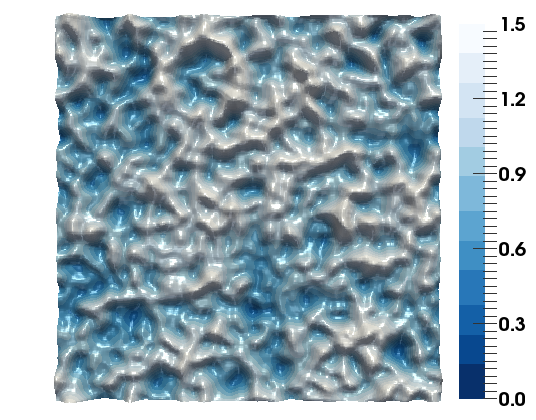}    \hfill \picA{$\varrho_{\rm e}$}{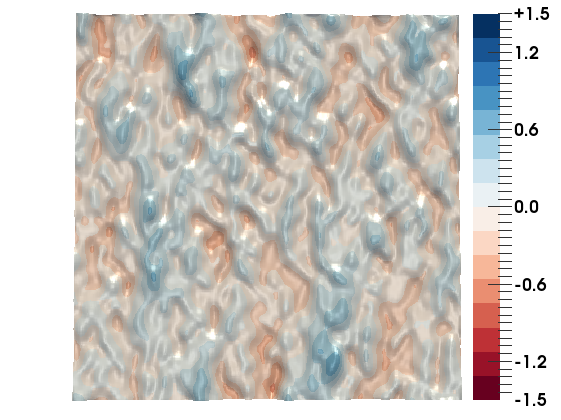}\hfill \picA{$|\Brho|$}{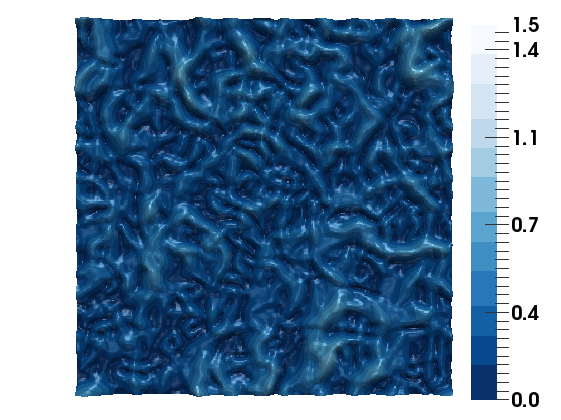}\\
\picA{$\qt/\rhot$}{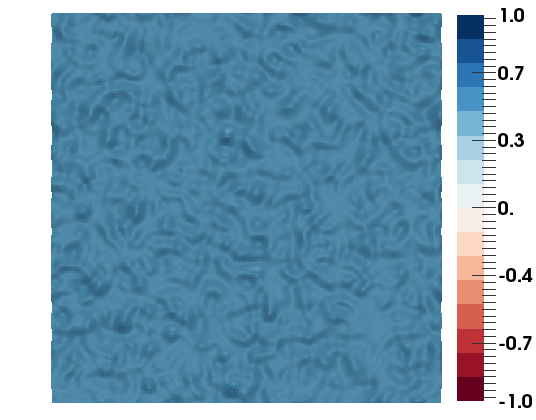}\hfill\; \picA{$\gamma$}{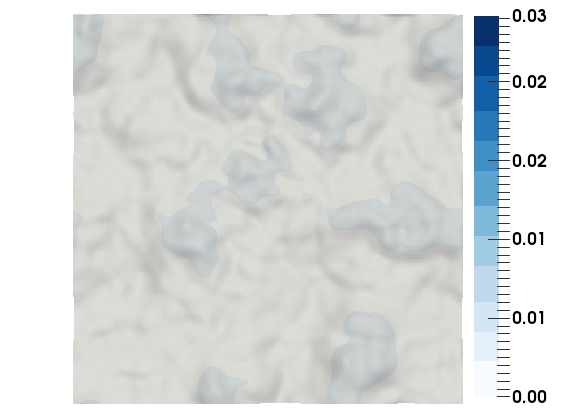}           \hfill \picA{$v$}{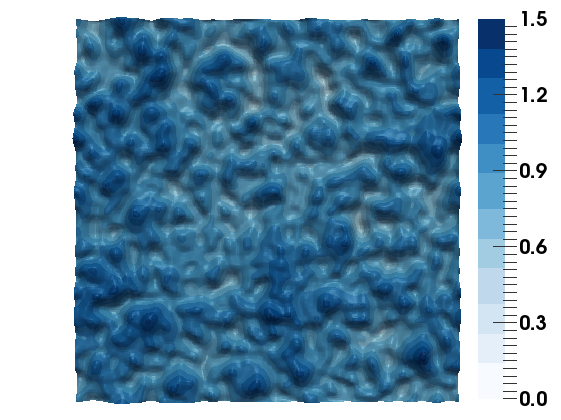}\\
\centering\vspace{3mm}
{\bf Stationary State}\\[-0.6em]
\picA{$\rhot$}{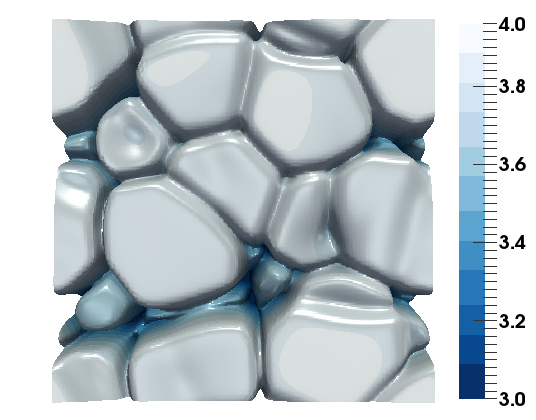}    \hfill \picA{$\varrho_{\rm e}$}{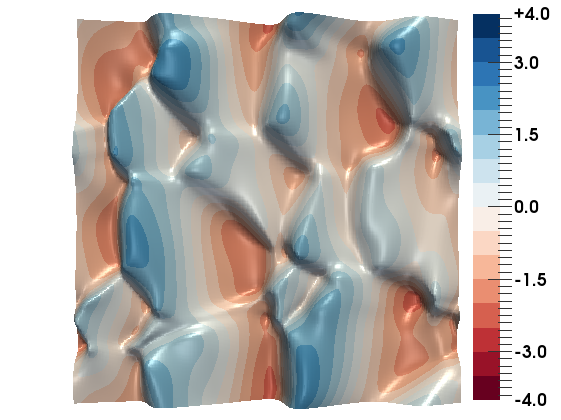}\hfill \picA{$|\Brho|$}{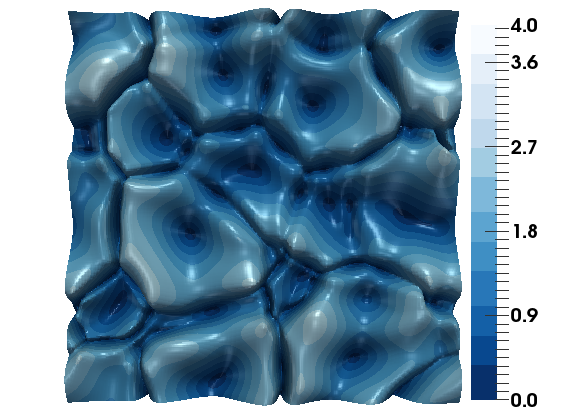}\\
\picA{$\qt/\rhot$}{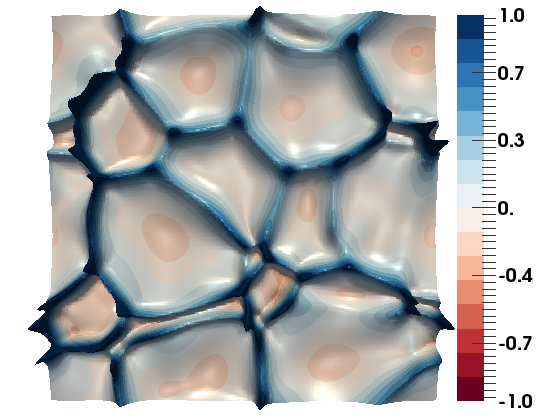}\hfill\quad \picA{$\gamma$}{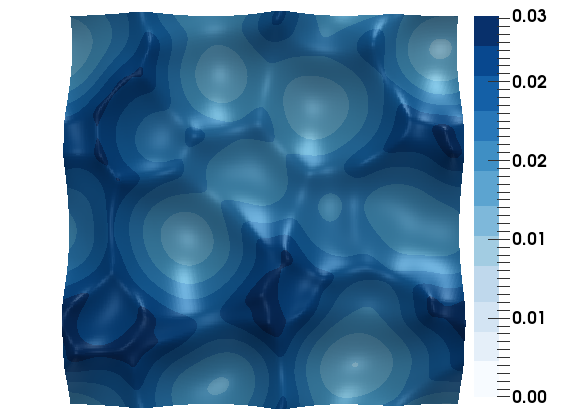}           \hfill \picA{$v$}{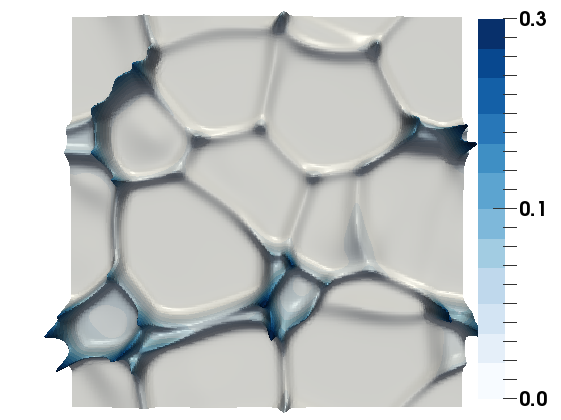}\\
\caption{Evolution of dimensionless CDD field variables, for parameters see Table~\ref{table:parameter}; all fields are represented in a projection on the slip plane, instead of \qt we show the more intuitive curvature \qt/\rhot\ which is the inverse of the local dislocation curvature radius, $\varrho_{\rm e}$ denotes the edge component of the GND density vector. Surfaces are plotted as elevated to visualize the fluctuations and are not up to scale. Top: initial field values as obtained from a random distribution of 50 dislocation loops, bottom: final stationary state; note that the scales have changed between top and bottom graphs}
\label{fig:pattern}
\end{figure}

\subsection{Simulation results}
\label{sec:results}
\begin{figure}[htbp]
\centering
%\subfloat[]{\includegraphics[width=0.4\textwidth]{figures/nloops}}
\subfloat[]{\includegraphics[width=0.495\textwidth]{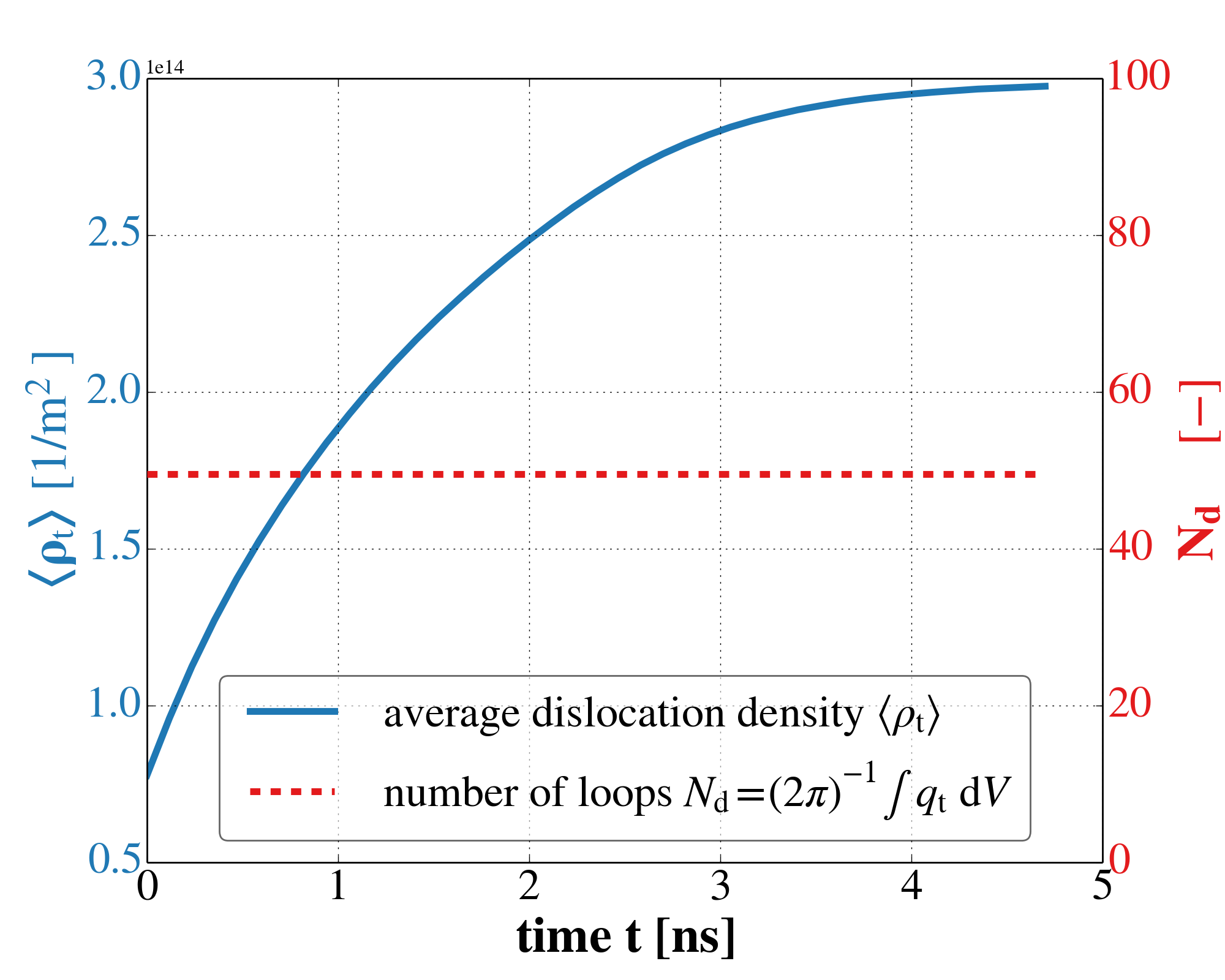}}
\hfill
%\subfloat[]{\includegraphics[width=0.4\textwidth]{figures/flux.png}}
\subfloat[]{\includegraphics[width=0.495\textwidth]{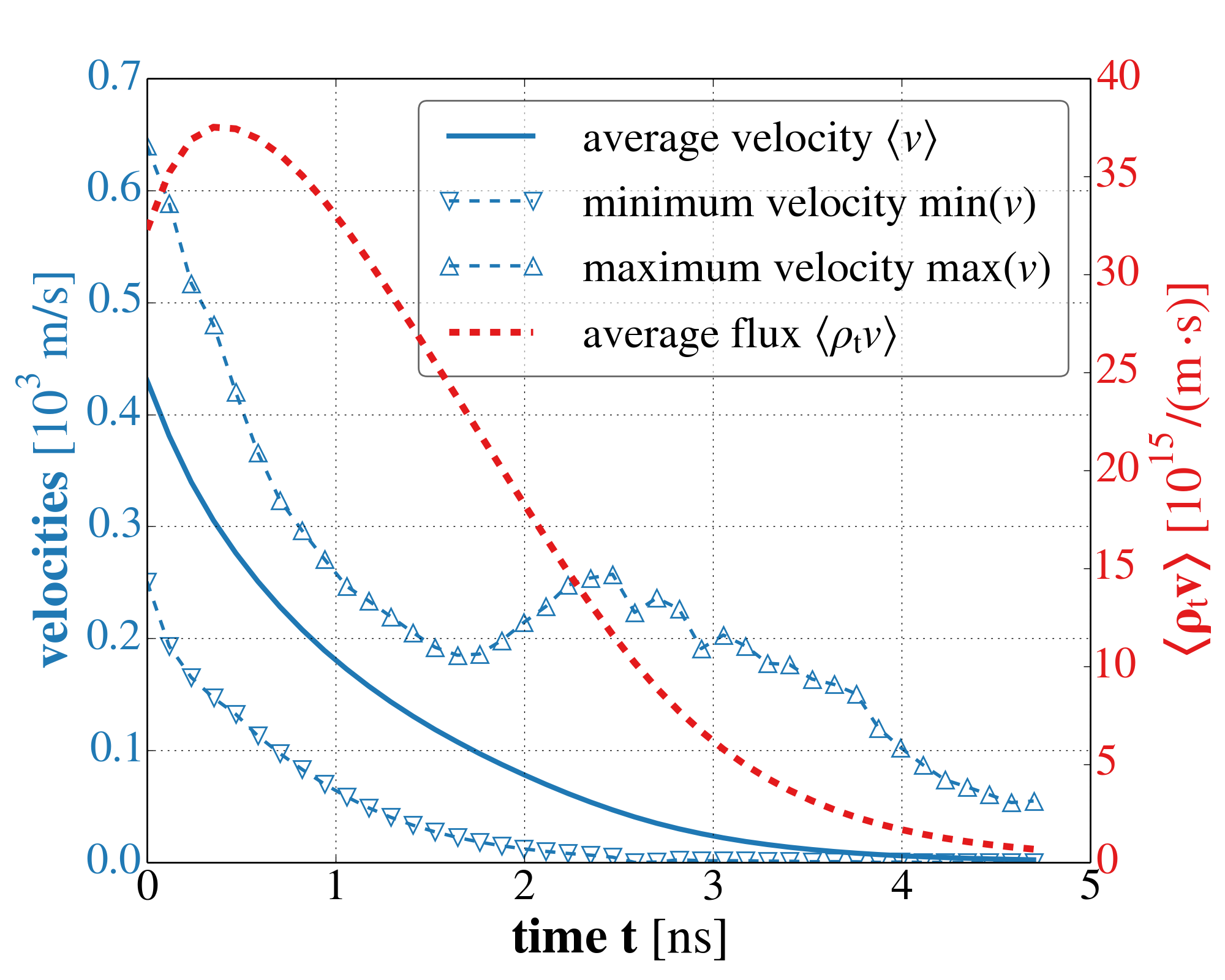}}
\caption{Evolution of average dislocation quantities in physical units: (a) average dislocation density $\langle \rhot \rangle$ and number of dislocation loops $N_d$; (b) average velocity $\langle v\rangle$, maximum and minimum velocity, and average dislocation flux $\langle\rhot v\rangle$ (mean strain rate).}
\label{fig:nloops}
\end{figure}
\begin{figure}[htbp]
\centering
\includegraphics[width=0.55\textwidth]{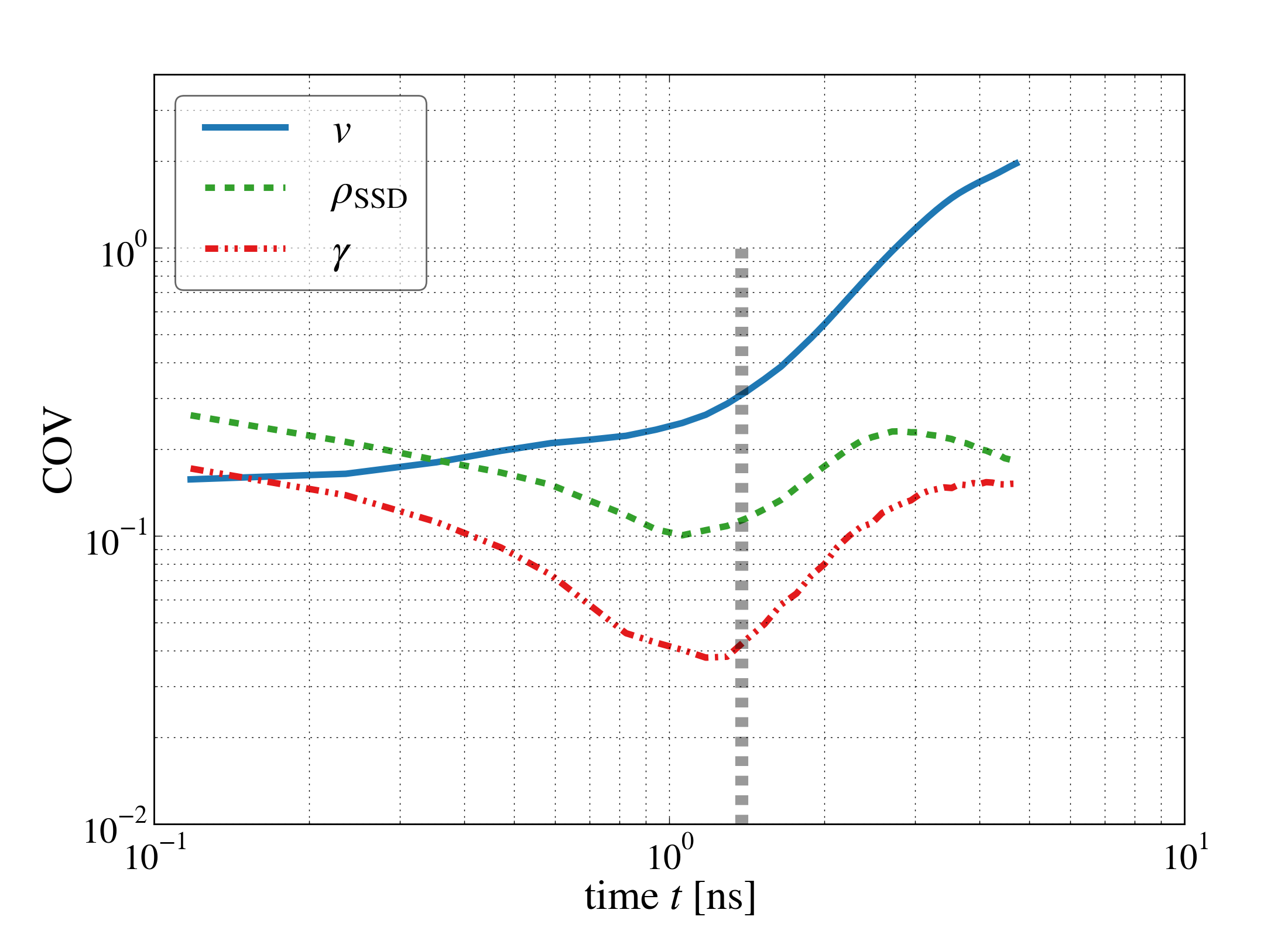}
\caption{Evolution of fluctuations of strain $\gamma$, statistically stored dislocation density $\rho_{\rm SSD} = \rhot - |\Brho|$, and dislocation velocity $v$ as characterized by the respective coefficients of variation ($\rm COV$); the dashed line indicates the point where the instability condition ${\cal H} = 0.725$ is first met.}
\label{fig:fluctuations}
\end{figure}

We first look at statistically averaged quantities (volume averages) before we then analyze the system in terms of microstructural patterning.

The mean dislocation density $\langle\rhot\rangle$ and loop number $N_{\rm d}$  are shown in \figref{fig:nloops} as functions of time. We observe that initially the dislocation density quickly increases until it saturates at a value that is about four times the initial value. At the same time the total number of dislocation loops $N_{\rm d}$ stays exactly constant. This is expected since loop creation by Frank-Read sources and loop merging by mutual annihilation of segments from different loops are not accounted for in the present simple model.  Accordingly, $\qt$ is a conserved quantity -- a fact which is immediately obvious from \eqref{eq:dqtdt}. The fact that $N_{\rm d}$ stays perfectly constant in our simulation indicates that the numerical scheme performs well.

In order to understand the saturation of the dislocation density, we note that the system can become stationary only by reducing the dislocation flux $\rhot v$ -- and thus the strain rate -- everywhere to zero. This is possible if either the Taylor stress balances the external stress, or if the local dislocation density vanishes. This observation agrees with the fact that, in the dislocation-rich regions, the local dislocation density approaches the saturation density given by $\rho\superscr{sat}= [\tauext/(aGb)]^2=2.99\cdot 10^{14}\,$m$^{-2}$ where the Taylor stress equals the applied external stress. 
The dynamics of the system is further illustrated in the right graph of \figref{fig:nloops} which shows the average dislocation flux (the total strain rate) and the mean, maximum and minimum dislocation velocity. Initially, we see an increase in dislocation flux which is driven by the increase in dislocation density. This is followed by a decrease in dislocation flux caused by the decrease in dislocation velocity due to strain hardening. Even in the final state, there remain regions of non-zero dislocation velocity which, however, no longer produce strain since they are depleted of dislocations.

\Figref{fig:pattern}, bottom, shows the CDD field variables in the stationary state; the supplementary movie additionally shows the time evolution of the total density. \Figref{fig:fluctuations} shows the concomitant evolution of the fluctuations of strain, dislocation density and dislocation velocity as characterized by the respective coefficients of variation ($\rm COV$, defined as the ratio between variance $\sigma$ and mean of the respective variable). In the initial period (cf. movie), the system is stable since the applied stress is significantly higher than the Taylor stress (${\cal H} < {\cal H}_{\rm c} = 6/7$). Accordingly, fluctuations of strain or dislocation density are initially smoothed out as seen from  \figref{fig:fluctuations}. However, the increase in dislocation density drives the system towards the critical state which is reached, for the simulated parameters, at a density of about $2.2 \times 10^{14}$ m$^{-2}$. This point in time is marked by the dashed line in \figref{fig:fluctuations}. The ensuing instability is characterized by an increase in the fluctuations of strain and dislocation density, and an accelerated growth of the fluctuations in dislocation velocity. Once the critical state is reached, persistent dislocation density 'blobs' emerge from the fluctuating density distribution while the channels between these blobs become depleted of dislocations. In these channels, dislocations move rapidly (the maximum dislocation velocity increases, see \figref{fig:nloops}) and curved line segments/loops expand and create additional line length which is deposited in the blobs. The emergent pattern is characterized by the following features (see \figref{fig:pattern}, bottom): 
\begin{enumerate}
\item 
Inside the dislocation-rich patches dislocations are to a significant part 'statistically stored', i.e. the 'geometrically necessary dislocation' (GND) density $|\Brho|$ is small.
\item 
Dislocations that form the boundaries of these dense patches are aligned to each other and hence become GNDs.
\item 
The 'channels' between the dislocation rich regions exhibit the largest plastic strain. Regions with high density, on the other hand, exhibit a reduced plastic strain.
\item
The curvature is reduced inside the dislocation-dense patches but is high in the dislocation depleted 'channels'. The dislocation dense patches act like hard inclusions which constrain dislocation motion. Particularly high curvature is observed in spots where "corners" of the dense patches  force the dislocations to assume strongly bent configurations. 
\item
The emerging pattern morphology does not depend on the initial values: a smaller initial density is compensated by a longer simulation time (more multiplication due to loop expansion) leading to the same final density and morphology. Simulations with different type of initial conditions (random fields vs. smeared-out loops) also produce the same final morphology. 
\item
The emergent pattern morphology does not depend on the numerical discretization (as long as the mesh is small enough to be able to represent the patterns shape); the pattern borders in general do not coincide with the finite element faces.
\item 
Patterns formed in different simulation runs with the same shear stress exhibit approximately the same characteristic length.
\end{enumerate}
The last observation is an auspicious feature of all our simulations - regardless of the chosen initial values, system size or material parameter. We analyze this behavior in the following section in terms of the 'law of similitude'.

\subsection*{Pattern evolution under changing stress and the principle of similitude}
In \figref{fig:pattern_sizes}(a) we show density patterns in simulations where we increase the external stress in a step-wise manner, such that each stress step is followed by a relaxation step until a quasi-stationary configuration is reached before the next stress increment. We record the saturation density  $\rho^\mathrm{sat}$ after each relaxation step.  The pattern morphology after relaxation remains essentially unchanged during the step sequence, however, the pattern size decreases with increasing stress while the dislocation density increases. To quantify this, we have determined the characteristic length of the patterns. This was done by use of the open source image analysis software \texttt{ImageJ} \cite{ImageJ}, which allows to determine the total number $n$, area fraction $A_i$ and perimeter $S_i$ of each density feature $i$ of the density distribution in a semi-automatic way. The characteristic length  is then obtained as the average value $\Lambda=\langle{4A_i/S_i}\rangle$ of the characteristic sizes of all relevant 'density blobs'. These data are plotted in \figref{fig:pattern_sizes}(b) together with the saturation density  $\rho^\mathrm{sat}$. \Figref{fig:pattern_sizes}(a) shows examples of the pattern morphology for the largest and smallest stresses.
\begin{figure}[htbp]
\centering
\subfloat[Plot (A) corresponds to a shear stress $\tauext=170\,\rm MPa$, the plot (B) is at $\tauext=560\,\rm MPa$. ]{
\includegraphics[height=0.45\textwidth]{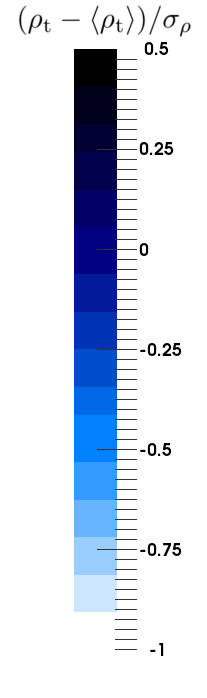}\quad\;
\includegraphics[height=0.46\textwidth]{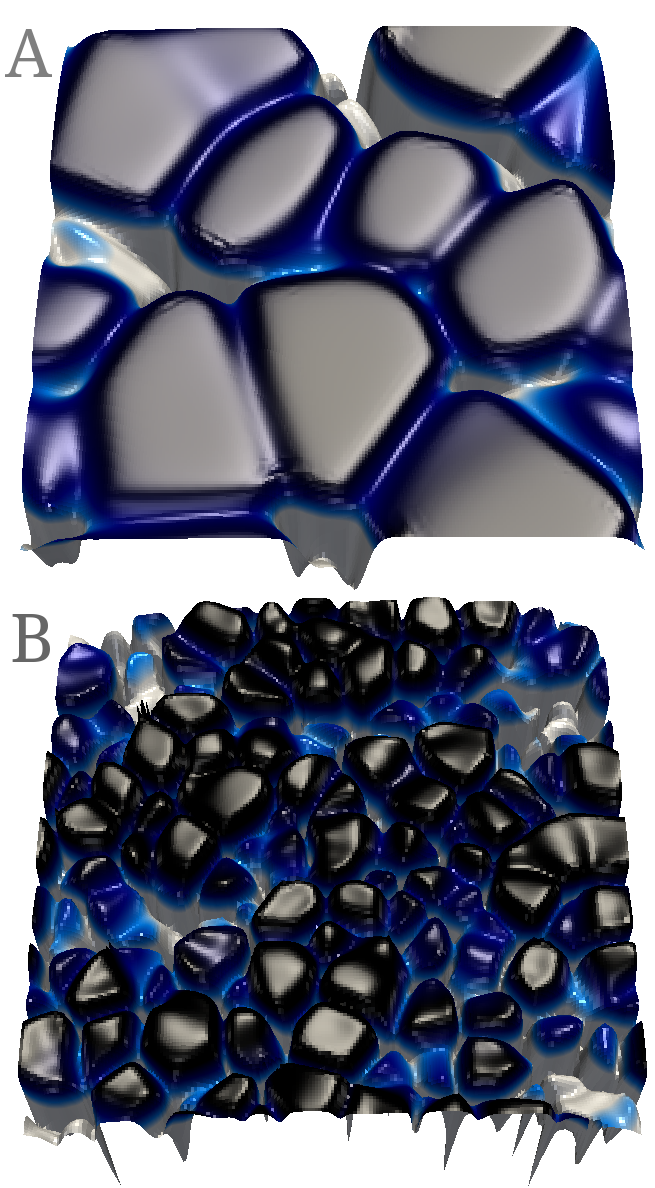}}
\hspace{6mm}
\subfloat[ %Stress vs. saturation dislocation density (blue triangles) and stress vs. patterning length (red diamonds). 
The two lines were fitted with constants $a = 0.3$ and $D=9.5$.]{\includegraphics[viewport=30 -30 725 550, clip, width=0.515\textwidth]{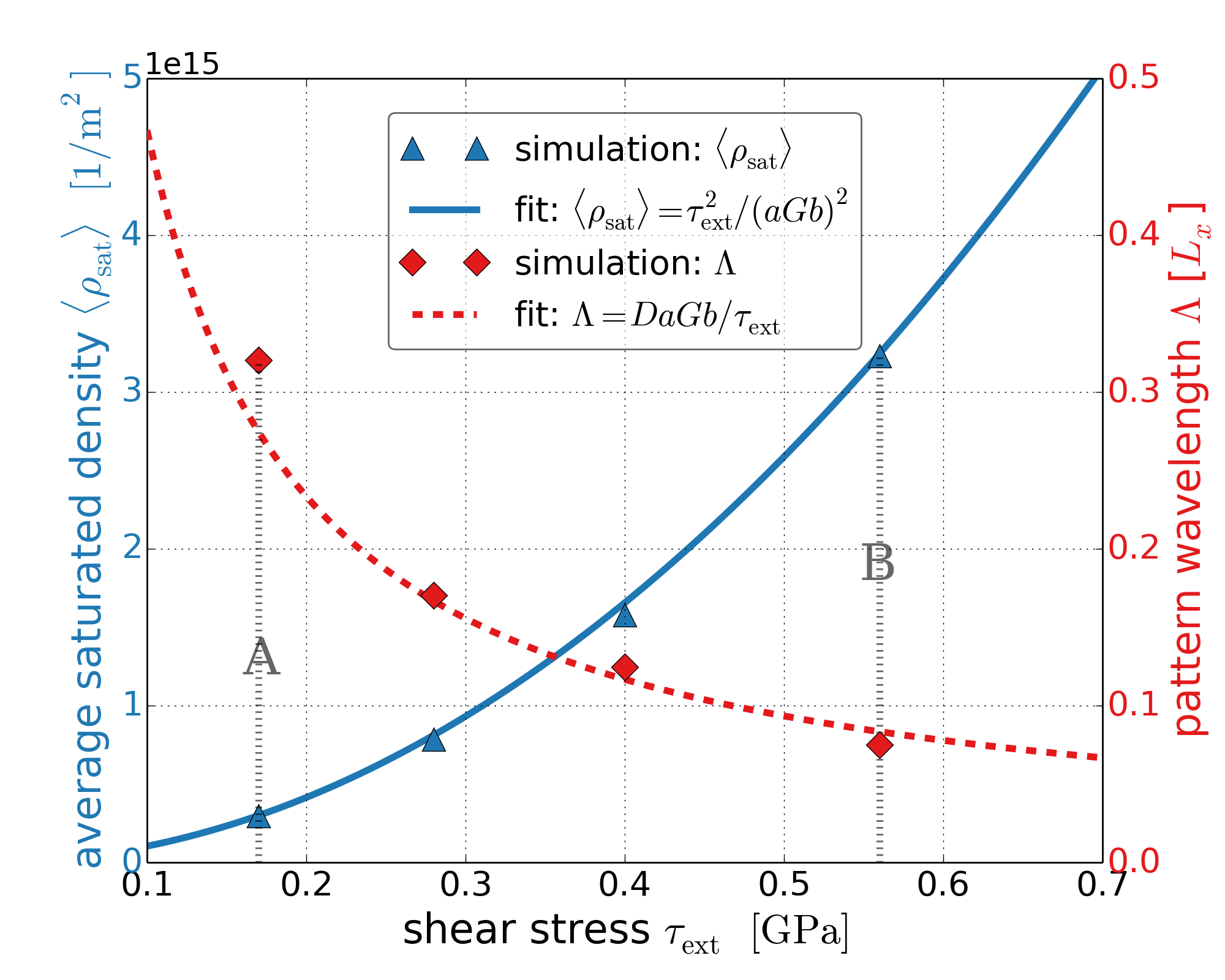}}\hfill
\caption{Relations between stress $\tauext$, patterning length $\Lambda$, and saturation dislocation density $\langle\rhot\rangle$ (right) and evolution of dislocation density patterns with increasing stress $\taue$ (left).}
\label{fig:pattern_sizes}
\end{figure}
Fitting the saturation density data by  $\langle\rho\superscr{sat}\rangle= [\tauext/(aGb)]^2$ with $a=0.3$ and the measured pattern wave length by  $\Lambda=DaGb/\tauext$ with $D=9.5$ we obtain good agreement for all data; thus the patterns are consistent with the similitude principle.

\section{Discussion and Conclusions}

We have formulated a model that describes the emergence of heterogeneous dislocation patterns that are consistent with the similitude principle. Linear stability analysis has demonstrated that the only requirement for the emergence of such patterns is the presence of short-range dislocation interactions as described by the Taylor relationship. Long-range internal stresses, on the other hand, are irrelevant for the occurrence of patterning -- although they may influence the pattern morphology by suppressing long wavelengths. 

The critical condition formulated in \secref{sec:analysis}, $0.725 < {\cal H} < 1$, implies a lower and an upper boundary to the range of stresses where patterning can occur. If $\taue$ is too small ($\taue < \tauy$), neither plastic flow nor dislocation patterning take place. This may appear trivial, but it formally demonstrates that the present model is inconsistent with the idea of quasi-equilibrium patterning due to energy minimization in the absence of externally driven plastic flow. If $\taue$ is too large, i.e, if the deformation rate is too high or, more generally speaking, the rate-dependent contribution to the flow stress becomes too large, then patterning is also suppressed. This observation simply means that the flow stress must be predominantly (to at least $72.5$ \%) attributed to dislocation interactions, and not to interactions with the crystal lattice (rate-dependent drag component). Qualitatively this is consistent with the fact that heterogeneous dislocation patterns are almost generically observed in fcc metals, but not in bcc metals at low temperatures where the flow stress is controlled by the interaction of dislocations with the lattice. 

\Insert{
A number of DDD simulations reported in the literature find that, in order to produce effective dislocation entanglement, the presence of cross slip is essential (see e.g. \cite{Madec2002_Scripta, Hussein2015_ActaMater85}). The same models show little or no significant increase in dislocation density with strain (work hardening) in the absence of cross slip. If, as we perceive it, dislocation patterning is essentially a corollary to work hardening, and DDD simulations require cross slip to produce hardening, then cross slip is in these simulations a prerequisite for dislocation patterning. We leave it open whether the same is true in real specimens since we think the role (or not) of cross slip in hardening, notably in multi-slip conditions, needs further investigation. 
}

It is interesting to compare the critical condition found in  the present case with the results of Zaiser \& Sandfeld \cite{Zaiser2014_MSMSE} who carried out a similar analysis for a system of straight parallel dislocations. In that case, the instability criterion has a similar structure. In fact, the mechanism underlying the instability is in both cases the same: The dislocation density evolution equations have the structure of conservation equations. Therefore, in regions of reduced dislocation velocity, the dislocation density increases and vice versa. Due to strain hardening, a local increase in dislocation density in turn reduces the local dislocation velocity, leading to instability. We thus find that patterning is a direct consequence of the short-range dislocation interactions which control strain hardening and of the kinematics of dislocations. No other ingredients are needed. 

What about dislocation multiplication? The present model naturally accounts for the increase of dislocation line length due to loop expansion. If we compare the resulting instability condition, $0.725 < {\cal H} < 1$, with the counterpart derived for straight parallel dislocations by Zaiser \& Sandfeld \cite{Zaiser2014_MSMSE}, which can be written as $2/3 < {\cal H} < 1$, we find that in the latter case the unstable band is more wide. The difference is due to the fact that in the present model which considers curved dislocations, regions of low dislocation density develop increased curvature and thus dislocation multiplication localizes in the dislocation-depleted regions - a fact which is well known from in-situ observations of dislocation motion in heterogeneous dislocation patterns. At the same time, dislocation multiplication is reduced in dislocation dense regions and this effect tends to counter-act the instability.   

\Insert{On a 'macroscopic' scale (i.e. the system scale) the average flow stress at saturation is slightly lower than in the case of a homogeneous dislocation distribution (\figref{fig:pattern_sizes}b). This is caused by the much lower flow stress in the nearly depleted 'channels' between the blobs and has been noted already by H. Mughrabi in \cite{Mughrabi1988_RevPhysAppl23}.}

Simulations of the model with increasing stress demonstrate that the patterns evolve in a manner that is consistent with the principle of similitude, i.e., the pattern wavelength is proportional to the mean dislocation spacing in the fully developed pattern, with a proportionality constant $D \approx 10$, and decreases in inverse proportion with the stress. All these features are consistent with observations. 

At the same time, it is important to emphasize the very substantial limitations of the presented model. With this study we do \emph{not} aim at realistically reproducing dislocation patterns observed in any real deformation experiment. Our aim is instead very much on the conceptual level: We want to present a minimal model which can be considered the skeleton of any realistic patterning model, while leaving it to future work to put in the realistic details. More realistic models will need to account for: 
\begin{itemize}
\item
Boundary conditions typical of real deformation experiments. We chose rather artificial boundary conditions to represent, for the purpose of analysis, the possibility of a situation without long-range internal stresses (case 1), or with only long-range stresses (case 2). In reality, of course, both features will always be present. 
\item \Insert{In this work we only considered bulk-like behavior. Finite-size effects as e.g. observed in \cite{Yu2014_PhilMag94,Hussein2015_ActaMater85} for single crystal pillars will play a role as soon as the pattern wavelengths approach the system size. Given that dislocations in our simulations generally have travelled about 1 wavelength in distance before they get frozen in the patterns it is very likely that the system needs to be significantly bigger than the pattern size, and that in small systems dislocation loss through the surfaces will inhibit patterning.}
\item
Though heterogeneous dislocation structures do form even in single slip \Insert{(i.e. in hardening stage I)}, situations with multiple active slip systems  are much more important. Accordingly, one needs to account for latent hardening terms coupling the dislocation densities on different slip systems in the expression for the short-range interaction stress, and for a more complex structure of the long-range internal stress field.
\item
The present model accounts for strain hardening in terms of the dislocation density increase that arises from loop expansion. However, in present form it does not represent a realistic hardening model: the reason is that according to \eqref{eq:dqtdt} the number of dislocation loops is a conserved quantity. This leads to a parabola-shaped hardening curve, $\tauy \propto \sqrt{\gamma}$, which is not found in any real material. The reason is simply that in real materials the loop number increases by the action of Frank-Read sources, whereas at large strains and/or dislocation densities also the merging of loops by annihilation of encountering segments needs to be taken into account. All these features, while needed for obtaining realistic hardening curves, are however incidental to the patterning mechanism. 
\item
In a realistic model, additional effects such as the influence of dislocation curvature/line tension on the dislocation velocity may need to be taken into account in an explicit manner. 
\end{itemize}
As a consequence of these limitations, the blob-like pattern morphology observed in the present model differs from the morphology of dislocation patterns commonly observed in real samples. Other models proposed in the literature (see \secref{intro}) produce much nicer patterns - but they cannot be shown to be in agreement with the similitude principle or to be based upon a realistic description of dislocation interactions. We are therefore convinced that the present approach, once the necessary model adjustments have been made, can serve as a robust framework for future investigation of dislocation patterning phenomena in a wide range of deformation settings. 

\section*{Acknowledgments}
The authors would like to thank Ha\"{e}l Mughrabi for stimulating discussions. S.~S. gratefully acknowledges financial support from the Deutsche Forschungsgemeinschaft
(DFG) through Research Unit FOR1650 'Dislocation-based Plasticity' (DFG grant No
Sa2292/1-1) and the European m-era.net project 'FASS' (DFG grant No Sa2292/2).

\section*{Bibliography}
\bibliographystyle{unsrt}
\bibliography{literature_sandfeld}

\end{document}